\documentclass[aps,prl,superscriptaddress,twocolumn]{revtex4}
\usepackage{bm}
\usepackage{graphicx,amsmath,latexsym,amssymb}
\usepackage{color}
\usepackage{dcolumn}

\begin{document}
\title{Quasi-resonant van der Waals interaction between non-identical atoms}
\author{M. Donaire} 
\email{donaire@lkb.upmc.fr, mad37ster@gmail.com}
\affiliation{Laboratoire Kastler Brossel, UPMC-Sorbonne Universit\'es, CNRS, ENS-PSL Research University,  Coll\`{e}ge de France, 4, place Jussieu, F-75252 Paris, France}
\author{R. Gu\'erout}
\affiliation{Laboratoire Kastler Brossel, UPMC-Sorbonne Universit\'es, CNRS, ENS-PSL Research University,  Coll\`{e}ge de France, 4, place Jussieu, F-75252 Paris, France}
\author{A. Lambrecht}
\affiliation{Laboratoire Kastler Brossel, UPMC-Sorbonne Universit\'es, CNRS, ENS-PSL Research University,  Coll\`{e}ge de France, 4, place Jussieu, F-75252 Paris, France}

\begin{abstract}
We present a time-dependent quantum calculation of the van der Waals interaction between a pair of dissimilar atoms, one of which is initially excited
while the other one is in its ground state. For small detuning, the interaction  is predominantly mediated  at all distances by the exchange of doubly resonant photons between the two atoms.
We find that it presents both temporal and spatial oscillations. 
Spatially oscillating terms depend on the 
resonant frequencies of both atoms, while the frequency of the time oscillations is given by their detuning.
We analyse the physical content of our findings and discuss to what extent previous conflicting stationary approaches provide compatible results. A proper account of causality is found essential in order to obtain the correct result.
\end{abstract}
\maketitle

Dispersion forces between neutral atoms are often interpreted as a result of the quantum fluctuation of both the electromagnetic (EM) field and the atomic charges \cite{Milonnibook,Cohenbook}.  A prominent example of those are van der Waals (vdW) forces acting between neutral atoms and molecules, which are important in  atomic and molecular interferometry where they influence the measured interference pattern \cite{Arndt2012}. In Quantum Information, vdW forces between Rydberg atoms produce a Rydberg blockade which may be exploited to realize quantum gates \cite{SaffmannRMP2010}.  In biophysical and chemical processes vdW forces are known to play a crucial role for the stability and assembling of molecules \cite{Israel}.\\
\indent At zero temperature, two atoms in their ground states undergo a series of virtual transitions to upper levels. It is the coupling of the charges of each atom to the quantum EM field that induces the correlation between their transient dipole moments, giving rise to a non-vanishing vdW interaction. The lifetime of the virtual atomic transitions is very short in comparison to ordinary observation times and thus, the use of stationary quantum perturbation theory is well justified for the calculation of this interaction \cite{Craigbook}. For short interatomic distances $r$ in comparison to the relevant transition wavelengths (non-retarded regime) the interaction scales 
as $r^{-6}$, while for large distances (retarded regime) it goes like $r^{-7}$ \cite{Milonnibook,Craigbook,London,Casimir-Polder1948,SafariPRA}.\\
\indent The situation is different for excited atoms. First, excited states are unstable and present finite lifetimes. 
This implies that, generically, the interaction between excited atoms must depend on time. Second, if any of the transitions from the excited to lower 
atomic levels is relevant to the interaction, the exchange of resonant photons between the atoms must be considered. The energy of the interaction
mediated by resonant photons is usually referred to as \emph{resonant van der Waals potential} in the literature \cite{Wylie,Buhmann,Gorza,PRAme}.  In the retarded regime the resonant potential overtakes by far  the non-resonant one.  It is in this regime that different approaches yield conflicting results concerning the spatial oscillations of the interaction \cite{Power1995,McLone,Power1965,Sherkunov,Henkel,SafariPRL,Passante,Berman2}. This long-standing problem is the main motivation of the present Letter.

 In the following, we address the 
time-dependent quantum computation of the interaction between two dissimilar atomic dipoles, one of which is excited. The excited atom is taken of
type $A$ while the atom in its ground state is considered of a different type $B$. Without loss of generality we approximate the atoms by two-level systems
of resonant frequencies $\omega_{A}$ and $\omega_{B}$ respectively, with respective linewidths $\Gamma_{A}$ and $\Gamma_{B}$.
Further, in order to ensure the perturbative nature of the calculation and to avoid resonant energy transfer we set the detuning 
$\Delta_{AB}\equiv\omega_{A}-\omega_{B}$ such that  $|\Delta_{AB}|>(\Gamma_{A}+\Gamma_{B})/2$ and $|\Delta_{AB}|\gg\langle W(T)\rangle/\hbar$, 
with $W(T)$ being the interaction Hamiltonian at the time of observation, $T$. Since the observation is made  for atom $A$ excited,
we must have $T\lesssim2\pi\Gamma_{A}^{-1}$. Lastly, we assume without much loss of generality $\Gamma_{A,B}<|\Delta_{AB}|\ll\omega_{A,B}$, which is easily met
by pairs of alkali atoms. We will refer to this condition as \emph{quasi-resonant}. We 
will see that it allows for a great reduction in the number of calculations and makes the resonant potential dominant at all distances.
We will show that the interaction energy oscillates both in time and in space. It contains time-independent terms which oscillate in space with frequency $2\omega_{A}/c$, and time-dependent terms which oscillate in time with frequency $\Delta_{AB}$ and in space with frequency $2\omega_{B}/c$. We compare our results to previous conflicting approaches and discuss in detail to which extent they provide compatible results.

We aim at computing  the EM energy of atom $A$ due to the presence of atom $B$. To this end we apply standard time-dependent quantum perturbative techniques in the electric dipole approximation \cite{Sakurai}.  At any given time $T$ the state of the two-atom-vacuum system can be written as $|\Psi(T)\rangle=\mathbb{U}(T)|\Psi(0)\rangle$, where the  state of the system at time 0 is $|\Psi(0)\rangle=|A_{+}\rangle\otimes|B_{-}\rangle\otimes|0_{\gamma}\rangle$. In this expression $(A,B)_{+,-}$ label the upper/lower internal states of the atoms $A$ and $B$ respectively and $|0_{\gamma}\rangle$ is the EM vacuum state. $\mathbb{U}(T)$ denotes the time evolution operator in the Schr\"odinger representation, 
\begin{equation}
\mathbb{U}(T)=\mathcal{T}\exp{}\Bigl\{-i\hbar^{-1}\int_{0}^{T}\textrm{d}t\Bigl[ H_{A}+H_{B}+H_{EM}+W\Bigr]\Bigr\}.\nonumber
\end{equation}
In this equation $H_{A}+H_{B}$ is the free Hamiltonian of the internal atomic states, $\hbar\omega_{A}|A_{+}\rangle\langle A_{+}|+\hbar\omega_{B}|B_{+}\rangle\langle B_{+}|$, while the Hamiltonian of the free EM field is $H_{EM}=\sum_{\mathbf{k},\mathbf{\epsilon}}\hbar\omega(a_{\mathbf{k},\mathbf{\epsilon}}a^{\dagger}_{\mathbf{k},\mathbf{\epsilon}}+1/2)$,
where $\omega=ck$ is the photon frequency, and the operators $a^{\dagger}_{\mathbf{k},\mathbf{\epsilon}}$ and $a_{\mathbf{k},\mathbf{\epsilon}}$ are the creation and annihilation operators of photons with momentum $\hbar\mathbf{k}$ and polarization $\mathbf{\epsilon}$ respectively. Finally, the interaction Hamiltonian reads $W=W_{A}+W_{B}$, with $W_{A,B}=-\mathbf{d}_{A,B}\cdot\mathbf{E}(\mathbf{R}_{A,B})$. In this expression $\mathbf{d}_{A,B}$ are the electric dipole operators of each atom and  $\mathbf{E}(\mathbf{R}_{A,B})$ is the electric field operator evaluated at the position of each atom, which   can be written in the usual manner as a sum over normal modes as \cite{Craigbook}
\begin{eqnarray}\label{AQ}
\mathbf{E}(\mathbf{R}_{A,B})&=&\sum_{\mathbf{k}}\mathbf{E}^{(-)}_{\mathbf{k}}(\mathbf{R}_{A,B})+\mathbf{E}^{(+)}_{\mathbf{k}}(\mathbf{R}_{A,B})\nonumber\\
&=&i\sum_{\mathbf{k},\mathbf{\epsilon}}\sqrt{\frac{\hbar ck}{2\mathcal{V}\epsilon_{0}}}
[\mathbf{\epsilon}a_{\mathbf{k}}e^{i\mathbf{k}\cdot\mathbf{R}_{A,B}}-\mathbf{\epsilon}^{*}a^{\dagger}_{\mathbf{k}}e^{-i\mathbf{k}\cdot\mathbf{R}_{A,B}}],\nonumber
\end{eqnarray}
where $\mathcal{V}$ is a generic volume and $\mathbf{E}^{(\mp)}_{\mathbf{k}}$ denote the annihilation/creation electric field operators of photons of momentum $\hbar\mathbf{k}$, respectively. While the internal atomic and EM degrees of freedom are quantum variables, the position vectors $\mathbf{R}_{A,B}$ are classical variables. We emphasize here that we do not make further simplifications to these potentials. In particular, we do not replace the EM response of any of the atoms by its ordinary polarizability, as it is the case in Ref.\cite{Passante}.

\begin{figure}[h]
\includegraphics[height=3.4cm,width=8.9cm,clip]{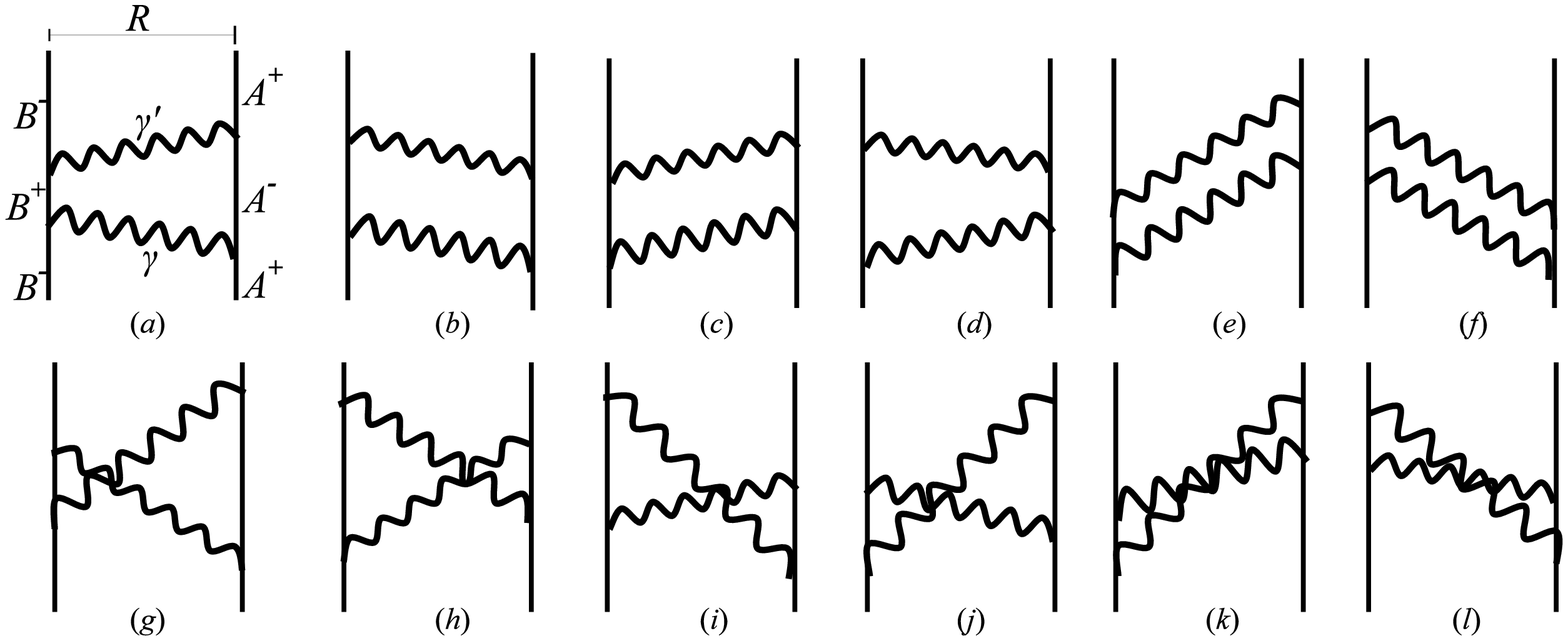}
\caption{Diagramatic representation of the twelve time-ordered processes which contribute to $\langle W_{A}(T) \rangle$  at the lowest order in $W$. The time variable runs along the vertical.}\label{fig1}
\end{figure}

Next, considering $W$ as a perturbation to the free Hamiltonians, the unperturbed time-evolution operator for atom and free photon states is $\mathbb{U}_{0}(t)=\exp{[-i\hbar^{-1}(H_{A}+H_{B}+H_{EM})t]}$. In order to make contact with a realistic setup, we imagine that atom $A$ starts being excited  at time $-\tau$ by a laser pulse of duration $\tau<|\Delta^{-1}_{AB}|$. This fixes our temporal resolution and implies that at  time $\simeq0$ the initial state $|\Psi(0)\rangle$ is well-defined within a time interval of the order of $\tau$. We are now ready to compute  the EM energy of atom $A$ due to the presence of atom $B$ at any time $T$ such that $0\lesssim T\lesssim2\pi\Gamma_{A}^{-1}$,
\begin{equation}
\langle W_{A}(T) \rangle=-\langle\Psi(0)|\mathbb{U}^{\dagger}(T)\mathbf{d}_{A}\cdot\mathbf{E}(\mathbf{R}_{A})\mathbb{U}(T)|\Psi(0)\rangle.\label{Force}
\end{equation}
The above expression admits an expansion in powers of  $W$ which can be developed out of the time-ordered exponential equation for $\mathbb{U}(T)$, $\mathbb{U}(T)=\mathbb{U}_{0}(T)\:\mathcal{T}\exp\int_{0}^{T}\mathbb{U}_{0}^{\dagger}(t)\:W\:\mathbb{U}_{0}(t)\textrm{d}t$. At leading order, Eq.(\ref{Force}) contains a series of terms of fourth order in $W$ where an electric field operator creates/annihilates a photon at time $T$ at the position of atom $A$. They correspond to the twelve well-known time-ordered diagrams of Fig.\ref{fig1} \cite{McLone,Craigbook}. In the time-dependent approach, each diagram contributes to $\langle W_{A}(T) \rangle$ with two terms in which $W_{A}$ is flanked by two $\mathbb{U}$-matrices which make the system evolve, in opposite time directions, from the initial state to two different states  at time $T$, which differ from one another in the state of atom $A$ and in the number of photons by one unit.  
In quasi-resonant conditions, the greatest contribution to $\langle W_{A}(T)\rangle$ comes from diagram $(a)$, in which two doubly resonant photons are exchanged one after the other. Doubly resonant photons are those emitted by one of the atoms in its upper level and absorbed by the other atom in its lower level, while for non-resonant photons the emission/absorpsion processes are inverted. Lastly, simply resonant photons are those emitted and absorbed by both atoms in either their upper or lower levels. In addition, the diagrams $(b)-(f)$  of Fig.\ref{fig1}, which contain both doubly resonant and non-resonant photons, provide terms which make it possible to extend the frequency integrals of diagram $(a)$ into the negative domain. Their contribution is indeed essential for establishing causality (cf. Ref.\cite{Berman}). All the other contributions from these and from the rest of diagrams are at the most of the order of $\Delta_{AB}/\omega_{A,B}$ times smaller and hence negligible. Putting everything together, transforming the sums over photon momenta into continuum integrals and imposing the causality condition $T>2R/c$ with $\mathbf{R}=\mathbf{R}_{B}-\mathbf{R}_{A}$, we find at leading order,
\begin{align}\label{laeq}
\langle W_{A}(T)\rangle&\simeq\frac{1}{2\hbar^{3}}\int_{-\infty}^{\infty}\frac{\mathcal{V}k^{2}\textrm{d}k}{(2\pi)^{3}}\int_{-\infty}^{\infty}\frac{\mathcal{V}k^{'2}\textrm{d}k'}{(2\pi)^{3}}\int_{0}^{4\pi}\textrm{d}\Omega\int_{0}^{4\pi}\textrm{d}\Omega'\nonumber\\
&\times\Bigl[i\langle\Psi(0)|\mathbb{U}_{0}(-T)|\Psi(0)\rangle\Theta(T-2R/c)\int_{0}^{T}\textrm{d}t\int_{0}^{t}\textrm{d}t'\nonumber\\
&\times\int_{0}^{t'}\textrm{d}t''
\langle\Psi(0)|\mathbf{d}_{A}\cdot\mathbf{E}_{\mathbf{k}'}^{(-)}(\mathbf{R}_{A})\mathbb{U}_{0}(T-t)\\
&\times\mathbf{d}_{B}\cdot\mathbf{E}_{\mathbf{k}'}^{(+)}(\mathbf{R}_{B})\mathbb{U}_{0}(t-t')\mathbf{d}_{B}\cdot\mathbf{E}_{\mathbf{k}}^{(-)}(\mathbf{R}_{B})\nonumber\\&\times\mathbb{U}_{0}(t'-t'')\mathbf{d}_{A}\cdot\mathbf{E}_{\mathbf{k}}^{(+)}(\mathbf{R}_{A})\mathbb{U}_{0}(t'')|\Psi(0)\rangle\Bigr]+[k\leftrightarrow k']^{\dagger}.\nonumber
\end{align} 
The time integrals of the time-evolution operators in Eq.(\ref{laeq}) yield a series of terms with  poles along the real axis,
\begin{align}
&\frac{c}{\Delta_{AB}(k-k_{A})(k'-k_{A})}-\frac{c\:\cos{(\Delta_{AB}T)}}{\Delta_{AB}(k-k_{B})(k'-k_{B})}\label{poles}\\
-&\frac{\cos{[(\omega-\omega_{A})T]}}{(k-k_{A})(k-k_{B})(k'-k)}+\frac{\cos{[(\omega'-\omega_{A})T]}}{(k'-k_{A})(k'-k_{B})(k'-k)}.\nonumber
\end{align} 
Further, the development of this expression contains terms in which both photons resonate either with the transition of atom $A$ or with the transition of atom $B$ only. This is a direct consequence of energy conservation. Upon integration in frequencies, the former terms are time-independent while the latter oscillate in time as $\sim\cos{\Delta_{AB}T}$. Important is the fact that only the first term in Eq.(\ref{poles}) arises in the stationary  approach \cite{McLone,Craigbook,SafariPRA}. However, the integration in frequencies of the third and fourth terms provides additional time-independent  contributions which are missing in the stationary approach. Lastly, replacing the time integrals in Eq.(\ref{laeq}) with the result (\ref{poles}) and integrating in orientations and frequencies, we obtain --see Appendix,
\begin{align}
\langle W_{A}(T)\rangle&=\frac{\mathcal{U}_{ijpq}}{R^{6}}[\beta^{ij}\beta^{pq}-k_{A}^{2}R^{2}(\beta^{ij}\beta^{pq}+2\alpha^{ij}\beta^{pq})\nonumber\\
&+k_{A}^{4}R^{4}\alpha^{ij}\alpha^{pq}]\cos{(2k_{A}R)}+\frac{2\mathcal{U}_{ijpq}}{R^{5}}k_{A}[\beta^{ij}\beta^{pq}\nonumber\\
&-k_{A}^{2}R^{2}\alpha^{ij}\beta^{pq}]\sin{(2k_{A}R)}\label{tdep}\\
&-\frac{\mathcal{U}_{ijpq}}{R^{6}}[\beta^{ij}\beta^{pq}-k_{B}^{2}R^{2}(\beta^{ij}\beta^{pq}+2\alpha^{ij}\beta^{pq})\nonumber\\
&+k_{B}^{4}R^{4}\alpha^{ij}\alpha^{pq}]\cos{(2k_{B}R+\Delta_{AB}T)}
-\frac{2\mathcal{U}_{ijpq}}{R^{5}}k_{B}\nonumber\\
&\times[\beta^{ij}\beta^{pq}-k_{B}^{2}R^{2}\alpha^{ij}\beta^{pq}]\sin{(2k_{B}R+\Delta_{AB}T)}\nonumber\\
&+\frac{\mathcal{U}_{ijpq}}{R^{6}}[1+...+(k_{A,B}R)^{4}]\mathcal{O}(\Delta_{AB}/\omega_{A,B})+...,\nonumber
\end{align}
where $\mathcal{U}_{ijpq}=\mu^{A}_{i}\mu^{A}_{q}\mu^{B}_{j}\mu^{B}_{p}/[(4\pi\epsilon_{0})^{2}\hbar\Delta_{AB}]$, 
$\mu^{A}=\langle A^{-}|\mathbf{d}_{A}|A^{+}\rangle$, $\mu^{B}=\langle B^{-}|\mathbf{d}_{B}|B^{+}\rangle$ and $\beta^{ij}=\delta^{ij}-3R^{i}R^{j}/R^{2}$, $\alpha^{ij}=\delta^{ij}-R^{i}R^{j}/R^{2}$. It is worth stressing that the Heaviside function in Eq.(\ref{laeq}) together with the time-order prescription do not only guarantee causality, but also determine univocally the contours of integration over frequencies in the complex plane when taking the principal value (see Appendix). The last term in Eq.(\ref{tdep}) indicates the order of the leading corrections to the dominant doubly resonant photon exchange terms of Eq.(\ref{laeq}) \cite{Comment}. 
As anticipated, the time-independent terms of Eq.(\ref{tdep}) oscillate only in space with frequency $2k_{A}$. On the contrary, the time-dependent terms oscillate in time with frequency $\Delta_{AB}$ and in space with frequency $2k_{B}$.  Only for large integration times, $\delta T\gg|\Delta_{AB}^{-1}|$, their time average vanishes. 
In the short time limit, $T\rightarrow 2R/c$,  $\langle W_{A}(T)\rangle$ vanishes identically at our order of approximation. This is a consequence of the fact that, in order to establish the interaction, it is necessary that the excitation be  transferred actually to atom $B$. For $T>R/c$, the probability of excitation of atom $B$ oscillates in time as $|\langle\Psi(T)|A_{-}\rangle\otimes|B_{+}\rangle\otimes|0_{\gamma}\rangle|^{2}\sim\sin^{2}{[\Delta_{AB}(R/c-T)]/2]}$, 
being maximum for the first time at $T=R/c+\pi|\Delta^{-1}_{AB}|$. Correspondingly, $\langle W_{A}(T)\rangle$ becomes maximum for the first time at $T=2R/c+\pi|\Delta_{AB}^{-1}|$. The lapse $R/c$ between these two times is the time for a photon to travel back from $\mathbf{R}_{B}$ to $\mathbf{R}_{A}$ after the excitation of atom $B$.

\begin{figure}[h]
\includegraphics[height=4.2cm,width=7.7cm,clip]{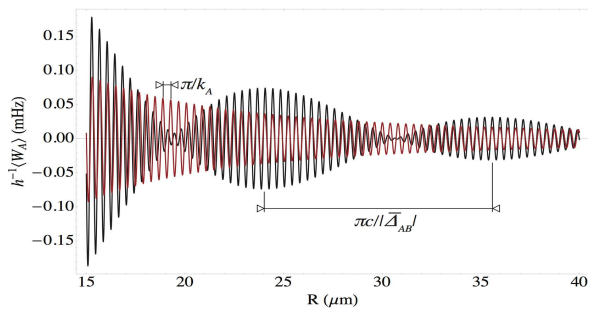}
\caption{Graphical representation of the interaction between a $^{87}$Rb atom in state $5P_{1/2}$ $(k_{A}=2\pi\:12578.95\:$cm$^{-1})$ and a $^{40}$K atom in its ground state,  for $k_{A}R\gg1$. The black line corresponds to a snapshot of the interaction at time $3.0\cdot10^{-12}\:$s, where the contributions of the  D1 and D2 transition lines of the  $^{40}$K atom ($k^{1}_{B}=2\pi\:12985.17\:$cm$^{-1}$ and  $k^{2}_{B}=2\pi\:13045.876\:$cm$^{-1}$ respectively) add up approximately in-phase. A long-range period $c\pi/|\bar{\Delta}_{AB}|$, with $\bar{\Delta}_{AB}/c=k_{A}-(k_{B}^{1}+k_{B}^{2})/2$, is identified. The red line corresponds to the time-independent result of the causal-adiabatic approximation. Average over dipole orientations has been taken.}\label{fig3}
\end{figure}

A long-standing debate exists in the literature concerning the spatial oscillations of the two-atom interaction in the retarded regime when one of the atoms is excited \cite{Power1995,McLone,Power1965,Sherkunov,Henkel,SafariPRL,Passante,Berman2}. The existence of spatial oscillations is indeed supported by experiments \cite{Wilson,Bushev}. According to our findings,  for $k_{A,B}R\gg1$ and $T>2R/c$, the interaction oscillates both in time and in space as
\begin{align}
\langle W_{A}(T)\rangle&\simeq\frac{\mathcal{U}_{ijpq}}{R^{2}}\alpha^{ij}\alpha^{pq}[k_{A}^{4}\cos{(2k_{A}R)}\label{FF}\\
&-k_{B}^{4}\cos{(2k_{B}R+\Delta_{AB}T)}]\nonumber\\
&\simeq\frac{-2\mathcal{U}_{ijpq}}{R^{2}}\alpha^{ij}\alpha^{pq}k_{A}^{4}\sin{[\Delta_{AB}(R/c-T/2)]}\nonumber\\
&\times\sin{[k_{A}(R+cT/2)+k_{B}(R-cT/2)]}.\nonumber
\end{align}
From the last expression we read that, at fixed time, the interaction is modulated by long-range oscillations of frequency $\Delta_{AB}/c$, while short-range oscillations take place at frequency $k_{A}+k_{B}$.  Also as a function of time the interaction is modulated by oscillations of frequency $\Delta_{AB}$. 
In Fig.\ref{fig3} we plot the energy of the interaction between two alkali atoms, one of $^{87}$Rb which is excited to the state $5P_{1/2}$  and another one of $^{40}$K which is in its ground state, in the retarded regime.

In contrast to our result, the stationary approach of Power and Thirunamachandran  in Ref.\cite{Power1995} predicts no oscillations for $\langle W_{A}\rangle$ in the far field. The key point in their calculation is the addition of small imaginary parts to the resonant frequency of atom $A$ in such a way that poles get shifted off the real axis. They used the prescription that a positive/negative imaginary part must be added for emitted/absorbed photons  in order to account for the finite linewidth of the excited atom. In particular, for  $\Delta_{AB}\ll\omega_{A,B}$ the dominant term in their stationary calculation is the first one in Eq.(\ref{poles}), but with the real poles shifted  as $[\Delta_{AB}(k-k_{A}-i\eta/c)(k'-k_{A}+i\eta/c)]^{-1}$, $\eta\rightarrow0^{+}$. 
After integrating in orientations an analogous equation to Eq.(\ref{laeq}) \cite{Craigbook}, they must have obtained for the energy in the far field limit, $k_{A}R\gg1$,
\begin{align}
-&\frac{\mathcal{U}_{ijpq}}{4\pi^{2}R^{2}}\alpha^{ij}\alpha^{pq}\int_{-\infty}^{+\infty}\textrm{d}k\int_{-\infty}^{+\infty}\textrm{d}k'k^{2}k^{'2}\label{Wad}\\
&\times\frac{e^{i(k+k')R}+e^{-i(k+k')R}-e^{i(k-k')R}-e^{-i(k-k')R}}{(k-k_{A}-i\eta/c)(k'-k_{A}+i\eta/c)},\nonumber
\end{align}
with $\eta\rightarrow0^{+}$. Since the pole in $k$ lies on the upper half of the complex plane and the pole in $k'$ lies on the lower half, 
the only nonvanishing contribution to the above integral comes from the term proportional to $e^{i(k-k')R}$. As the real part of the poles is in both cases $k_{A}$, taking the limit $\eta\rightarrow0^{+}$ the exponent vanishes after evaluating the residues and the integral yields the non-oscillating result $\frac{\mathcal{U}_{ijpq}}{R^{2}}k_{A}^{4}\alpha^{ij}\alpha^{pq}$.

In the previous stationary calculation of  
McLone and Power \cite{McLone} and Gomberoff \emph{et al.} \cite{Power1965} the poles in Eq.(\ref{Wad}) were not shifted. As a result, when taking the principal value of the integrals in Eq.(\ref{Wad}) with $\eta=0$, the four exponentials in the numerator contribute as $\sim 2\cos{2k_{A}R}+2=4\cos^{2}{k_{A}R}$ after adding up the residues, yielding the oscillating result  $\frac{\mathcal{U}_{ijpq}}{R^{2}}k_{A}^{4}\alpha^{ij}\alpha^{pq}\cos^{2}k_{A}R$.  As mentioned after Eq.(\ref{poles}), some time-independent terms are missing in the stationary calculation, which explains the discrepancy of this result with the time-independent component of ours in Eq.(\ref{FF}). In fact, the correct time-independent result can be obtained  by switching on adibatically the interaction potential, $W$, within the time-dependent causal approach. That is, by replacing $W_{A,B}(t)$ in Eq.(\ref{laeq}) with $W_{A,B}(t)\:e^{\eta t}$, $\eta\rightarrow0^{+}$, and extending the lower limit of the time integrals to $-\infty$--see Appendix. This results in an equation analogous to Eq.(\ref{Wad}) but with both poles shifted along the imaginary axis in the same direction an amount $i\eta/c$, which yields $\frac{\mathcal{U}_{ijpq}}{R^{2}}k_{A}^{4}\alpha^{ij}\alpha^{pq}\cos{2k_{A}R}$ for $k_{A}R\gg1$.

Recently, Safari and Karimpour have published a letter \cite{SafariPRL} where they claim to obtain for $k_{A}R\gg1$ the same oscillating behaviour as Gomberoff \emph{et al.} \cite{Power1965}. However, a straightforward comparison of Eq.(19) of Ref.\cite{SafariPRL} and Eqs.(14,26) of Ref. \cite{Power1965} reveals that this is indeed not the case. Whereas the result of the latter is the one outlined above, $\sim\cos^{2}k_{A}R$, the authors of the former have found $\sim\cos{2k_{A}R}$, despite the fact that both approaches are based on fourth order stationary perturbation theory. The origin of the discrepancy is in the algebraic manipulation inherited by the authors of Ref.\cite{SafariPRL} from Ref.\cite{SafariPRA}. In the Appendix B of Ref.\cite{SafariPRA} the authors have tried to express  the total contribution of the twelve diagrams of Fig.\ref{fig1} as a single frequency integral whose integrand is a function of the ordinary polarizabilities of the two atoms. In doing so by means of Eq.(B2) of Ref.\cite{
SafariPRA}, the authors have replaced effectively the denominator of Eq.(\ref{Wad}), which is a symmetric and separable function of $k$ and $ k'$ for $\eta=0$, by the expression $[\Delta_{AB}(k-k_{A})]^{-1}[1/(k'-k)+1/(k'+k)]$, which is neither symmetric nor separable. As a consequence, that replacement makes the frequency integrals depend arbitrarily on the order of integration. Next, integrating in $k'$ first and in $k$ later, one obtains $\frac{\mathcal{U}_{ijpq}}{R^{2}}k_{A}^{4}\alpha^{ij}\alpha^{pq}\cos{2k_{A}R}$, which agrees with Eq.(19) of Ref.\cite{SafariPRL} for $k_{A}R\gg1$, $\Delta_{AB}\ll\omega_{A,B}$, upon averaging in atomic orientations. Interestingly, this result equals the time-independent term of Eq.(\ref{FF}). However, this coincidence can only be accidental, since the above replacement and the subsequent prescription on the order of integration are neither connected to the time-dependent terms of Eq.(\ref{poles}) which cause the actual discrepancy with respect to the result of Refs.\cite{McLone,Power1965} nor to the causal-adiabatic approximation. 

\begin{figure}[h]
\includegraphics[height=2.2cm,width=8.2cm,clip]{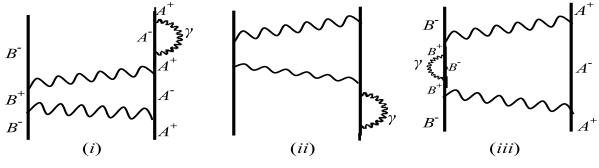}
\caption{Dominant diagrams at order $\mathcal{O}(W^{6})$ which incorporate  in the calculation of $\langle W_{A}(T)\rangle$ the effect of photon emission into free space. In $(i)$ and $(ii)$ the photon $\gamma$ is emitted into free space from atom $A$, whereas in $(iii)$ it is emitted from atom $B$.}\label{fig2}
\end{figure}

It is worth noting that while we have invoked the existence of finite lifetimes $\sim\Gamma^{-1}_{A,B}$ in order to impose physical constraints on the detuning $\Delta_{AB}$ and on the observation time $T$, no explicit reference to these quantities appear in our expression for $\langle W_{A}(T)\rangle$.  As a matter of fact, only the emission through the exchange of resonant photons between the two atoms has been implicitly accounted for in our calculation of $\langle W_{A}(T)\rangle$. 
 However, our calculation lacks the inclusion of the spontaneous emission of each atom into free space, whose rates are $k_{A,B}^{3}\mu_{A,B}^{2}/3\pi\epsilon_{0}\hbar$, respectively. The processes corresponding to the latter phenomenon are generally  unimportant in comparison to those depicted in Fig.\ref{fig1} since their leading contribution to $\langle W_{A}(T)\rangle$ is of order $\mathcal{O}(W^{6})\sim\mu_{A}^{4}\mu_{B}^{2},$ $\mu_{A}^{2}\mu_{B}^{4}$ --see Fig.\ref{fig2}. They might only be relevant for the case that the lifetimes are of the order of the temporal frequency of the interaction, $\Gamma_{A,B}\sim|\Delta_{AB}|$, but  they cannot affect in any case the  oscillatory behaviour found here for the terms of order $\mu_{A}^{2}\mu_{B}^{2}$. This argument opposes to the reasons given in Ref.\cite{Power1995} to add imaginary shifts to the real poles at $\mathcal{O}(W^{4})$. We finalize by mentioning that Berman has shown in a recent publication \cite{Berman2} how to introuduce spontaneous emission in a consistent manner in an adiabatic approximation. Assuming that atom $A$ is excited adiabatically with $\tau,\delta T\gg|\Delta_{AB}^{-1}|$, he has obtained the correct time-independent result.

In this Letter we have shown that the van der Waals interaction between two dissimilar atoms, one of which is initially prepared in an excited state, presents generically oscillations both in time and in space.
In quasi-resonant conditions the interaction is dominated at all distances by the exchange of doubly resonant photons between the two atoms. It is modulated in space by long-range oscillations of frequency $\Delta_{AB}/c$, while short-range
oscillations take place at frequency $k_{A}+k_{B}$. The time frequency is $\Delta_{AB}$, which determines the rate at which the excitation is transferred
to atom $B$. In the retarded regime the interaction takes the form of Eq.(\ref{FF}). 
Only for large integration times, $\delta T\gg|\Delta_{AB}^{-1}|$, that expression reduces to a time-independent term which oscillates in space with frequency $2k_{A}$. The latter is  also equivalent to the result of the causal-adiabatic approximations (see Appendix and Ref.\cite{Berman2}). It does not agree, however, with the result of stationary perturbation theory \cite{McLone,Power1965}. 

\acknowledgments
We thank Paul Berman and Marie-Pascale Gorza for fruitful discussions. 
Financial support from ANR-10-IDEX-0001-02-PSL and ANR-13-BS04--0003-02 is gratefully acknowledged.
\appendix

\section{Appendix: Integration of Eq.(\ref{laeq}) in the complex plane and the causal-adiabatic approximation}\label{Append}

In the first place, we note that all the time-evolution free-propagators in Eq.(2) are retarded,  $\mathbb{U}_{0}(t^{a}-t^{b})$ with $t^{a},t^{b}=0,t'',t',t,T$.  This is so because the time-order prescription on the time integrals implies $T\geq t'\geq t''\geq0$,  hence $t^{a}\geq t^{b}$ holds in all the $\mathbb{U}_{0}(t^{a}-t^{b})$ oparators of Eq.(2). It is the time-ordering  together with the Heaviside function $\Theta(T-2R/c)$ in the integrand of Eq.(\ref{laeq}) that guarantee causality.

Substituting in Eq.(\ref{laeq}) the expressions of the operators $\mathbf{E}^{(\pm)}(\mathbf{R}_{A,B})$ as sums over normal modes, and making use of the identities \cite{Craigbook} $\sum_{\mathbf{\epsilon}(\mathbf{k})}\epsilon_{i}(\mathbf{k})\epsilon^{*}_{j}(\mathbf{k})=\delta_{ij}-k_{i}k_{j}/k^{2}$ and
\begin{align}
&\int^{4\pi}_{0}\frac{\textrm{d}\Omega}{4\pi}(\delta_{ij}-k_{i}k_{j}/k^{2})e^{i\mathbf{k}\cdot\mathbf{R}}=
\alpha_{ij}\sin{kR}/kR\nonumber\\
&+\beta_{ij}\bigl[\cos{kR}/(kR)^{2}-\sin{kR}/(kR)^{3}\bigr],\label{SM1}
\end{align}
that we will denote by  $\mathcal{F}_{ij}(kR)$, we can write Eq.(\ref{laeq}) as
\begin{eqnarray}
\langle W_{A}(T)\rangle&\simeq&\frac{\alpha c^{3}}{8\pi^{3}\epsilon_{0}e^{2}}\mu^{A}_{i}\mu^{B}_{j}\mu^{B}_{p}\mu^{A}_{q}\int_{-\infty}^{+\infty}\textrm{d}k\:k^{3}\mathcal{F}^{ij}(kR)\nonumber\\&\times&\int_{-\infty}^{+\infty}\textrm{d}k'\:k'^{3}\mathcal{F}^{pq}(k'R)
\int_{0}^{T}\textrm{d}t\:\Theta(T-2R/c)\nonumber\\&\times&
\int_{0}^{t}\textrm{d}t'\int_{0}^{t'}\textrm{d}t''\Bigl[\bigl(i\:e^{i\omega_{A}T}e^{-i(T-t)\omega}\label{SM2}\\
&\times&e^{-i(t-t')\omega_{B}}e^{-i(t'-t'')\omega'}e^{-it''\omega_{A}}\bigr)+(\omega\leftrightarrow\omega')^{\dagger}\Bigr],\nonumber
\end{eqnarray}
where the integrand of the time-ordered integrals is the product of the retarded time-evolution free-propagators, $\alpha=e^2/4\pi\epsilon_{0}\hbar c$ is the fine-structure constant and $e$ is the electron charge. Next, integrating in time we end up with
\begin{eqnarray}
\langle W_{A}(T)\rangle&\simeq&\frac{\alpha}{4\pi^{3}\epsilon_{0}e^{2}}\mu^{A}_{i}\mu^{B}_{j}\mu^{B}_{p}\mu^{A}_{q}\textrm{P.V.}\int_{-\infty}^{+\infty}\textrm{d}k\:k^{3}\mathcal{F}^{ij}(kR)\nonumber\\
&\times&\int_{-\infty}^{+\infty}\textrm{d}k'\:k'^{3}\mathcal{F}^{pq}(k'R)\Theta(T-2R/c)\nonumber\\
&\times&\Bigr[
\frac{c/\Delta_{AB}}{(k-k_{A})(k'-k_{A})}-\frac{c/\Delta_{AB}\:\cos{(\Delta_{AB}T)}}{(k-k_{B})(k'-k_{B})}\nonumber\\
&-&\frac{\cos{[(\omega-\omega_{A})T]}}{(k-k_{A})(k-k_{B})(k'-k)}\nonumber\\&+&\frac{\cos{[(\omega'-\omega_{A})T]}}{(k'-k_{A})(k'-k_{B})(k'-k)}\Bigr],\label{SM3}
\end{eqnarray}
where straightforward application of perturbation  theory enforces to taking the principal value (P.V.) of the integral in Eq.(\ref{SM3}), as intermediate states with energies equal to that of the initial state must be removed from the sums (integrals in the continuum) in order to avoid zeros in the denominators,
\begin{align}
\sum_{k\neq k_{0}}\frac{\mathcal{F}(k)}{k-k_{0}}&\rightarrow\int_{-\infty}^{k_{0}-\eta/c}\frac{\mathcal{F}(k)}{k-k_{0}}\textrm{d}k+\int^{+\infty}_{k_{0}+\eta/c}\frac{\mathcal{F}(k)}{k-k_{0}}\textrm{d}k\nonumber\\
&=\textrm{P.V.}\int_{-\infty}^{+\infty}\frac{\mathcal{F}(k)}{k-k_{0}}\textrm{d}k\quad\textrm{as }\eta\rightarrow0^{+}.\label{SM4}
\end{align}
Having impossed causality by time-ordering the integrals of Eq.(\ref{SM2}) and multiplying the integrand by $\Theta(T-2R/c)$,  there is no need to add arbitrarily small imaginary parts $\pm i0^{+}$ to the real poles of the  integrand in Eq.(\ref{SM3}) in order to ensure that the propagators are retarded. Also, we have argued in the Letter that spontaneous emission enters at higher order, hence it cannot provide imaginary shifts $\pm i\Gamma_{A,B}/c$ to the real poles of Eq.(\ref{SM3}) either, in contrast to Refs.\cite{Power1995,Berman2}.  Lastly, the contours of integration  for each of the terms of the frequency integrals in the complex plane is univocally determined by the condition $T>2R/c$. Each term contains complex exponential factors of $k$ and $k'$ of the form $e^{ik\:f_{1}(R,T)}e^{ik'f_{2}(R,T)}$, with $f_{1,2}(R,T)$ being generic linear functions of $R$ and $T$.  That implies that the integration contours must be closed by infinitely large semi-circles either in the upper half plane for $f_{1,2}(R,T)>0$, or  in the lower half plane for $f_{1,2}(R,T)<0$, which is ultimately determined by the condition $T>2R/c$. It is worth mentioning that the integrand of Eq.(\ref{SM3}) is invariant under the exchange $k\leftrightarrow k'$, which implies that the total integral is independent of the order of integration, as expected 
\footnote{Obviously, it is understood that once a particular order of integration has been chosen, it must be applied to all the terms of the integrand which are not separable functions of $k$ and $k'$.}. Thus, $\langle W_{A}(T)\rangle$ is unambiguously given by Eq.(\ref{SM3}). The final result is given in the Eq.(4) of the Letter.

Finally, the causal-adiabatic approximation referred to in the conclusions of the Letter corresponds to adding up a factor $e^{\eta(t+t'+t'')}$ to the integrand of Eq.(\ref{SM2}), with $\eta\rightarrow0^{+}$, and to extending the lower limits of the time integrals to $-\infty$. After performing the  time integrals one obtains the $T$-independent terms of Eq.(\ref{SM3})  (excluding the Heaviside function) with a small imaginary part $- i\eta/c$  added to both real poles,
\begin{align}
&\frac{\alpha}{4\pi^{3}\epsilon_{0}e^{2}}\mu^{A}_{i}\mu^{B}_{j}\mu^{B}_{p}\mu^{A}_{q}\int_{-\infty}^{+\infty}\textrm{d}k\:k^{3}\mathcal{F}^{ij}(kR)\int_{-\infty}^{+\infty}\textrm{d}k'k'^{3}\nonumber\\&\times
\mathcal{F}^{pq}(k'R)\frac{c}{\Delta_{AB}(k-k_{A}-i\eta/c)(k'-k_{A}-i\eta/c)}.\label{SM5}
\end{align} 
We also note that  Eq.(6) in the Letter corresponds to the $T$-independent terms of Eq.(\ref{SM3}) of which only the terms proportional to $\alpha^{ij}$ and $\alpha^{pq}$ have been retained in the $\mathcal{F}$-functions, and in which small imaginary parts $\pm i\eta/c$ have been added to each real pole respectively.


\begin{thebibliography}{105}
\bibitem{Comment} Some of the diagrams of Fig.(\ref{fig1})  discarded  provide also terms  with poles in both
$k$ and $k'$. However, after integrated in frequencies, they yield contributions of the order of $\Delta_{AB}/(\omega_{A}+\omega_{B})$ times
smaller than the ones found here, and hence they are negligible. For instance, in diagram $(g)$ this is due to its lack of resonance with the transition
of atom $B$, since both photons are simply resonant.
\end{thebibliography}
\end{document}